\documentclass{aa}

\usepackage{graphicx}
\usepackage{txfonts}
\usepackage{natbib}

\begin{document}

   \title{Aromatic emission from the ionised mane of the Horsehead 
          nebula\thanks{This
          work is based on observations made with the Spitzer Space
          Telescope, which is operated by the Jet Propulsion Laboratory,
          California Institute of Technology under a contract with NASA.}}

   \subtitle{}
  
   \titlerunning{AIB from the ionised mane of the Horsehead 
          nebula}

   \author{M. Compi\`egne\inst{1}
           \and
            A. Abergel\inst{1}
           \and
            L. Verstraete\inst{1}
           \and
            W. T. Reach\inst{2}
           \and
            E. Habart\inst{1}  
           \and    
           J.D. Smith\inst{4}
           \and            
            F. Boulanger\inst{1}
           \and 
           C. Joblin\inst{3}
       }

   \offprints{M. Compi\`egne, \\
              email: {\bf Mathieu.Compiegne@ias.fr}}

   \institute{Institut d'Astrophysique Spatiale, UMR8617, CNRS, Universit\'e Paris-sud XI,
              b\^atiment 121, F-91405 Orsay Cedex, France
              \and
              Spitzer Science Center (SSC), California Institute of Technology, 
              1200 East California Boulevard, Pasadena, CA 91125
              \and
              Centre d'Etude Spatiale des Rayonnements, CNRS et Universit\'e Paul Sabatier-Toulouse 3,
              Observatoire Midi-Pyr\'en\'ees, 9 Avenue du Colonel Roche, 31028 Toulouse cedex 04, France
              \and
              Steward Observatory, University of Arizona, Tucson, AZ 85721 
                }

   \date{Received ; accepted }

 \abstract
{This work is conducted as part of the ``SPECPDR'' program
dedicated to the study of very small particles and chemistry in
photo-dissociation regions with the Spitzer Space Telescope (SST).}
{We study the evolution of the Aromatic
Infrared Bands (AIBs) emitters across the illuminated edge of the
Horsehead nebula and
especially their survival and properties in the HII region.}
{We present spectral mapping observations
taken with the Infrared Spectrograph (IRS)
at wavelengths 5.2-38\,$\mu$m. The spectra have a resolving power 
of $\rm{\lambda/\Delta \lambda}$\,=\,64\,-\,128 and show the main aromatic bands, H$_2$
rotational lines, ionised gas lines and continuum. The maps have an
angular resolution of 3.6-10.6\arcsec and allow us to study the
nebula, from the HII diffuse region in front
of the nebula to the inner dense region.}
{A strong AIB at 11.3\,$\mu$m is detected in
the HII region, relative to the other AIBs at 6.2, 7.7 and 8.6\,$\mu$m, 
and up to an angular separation of $\sim$\,20\,\arcsec (or 0.04\,pc)
from the ionisation front.  The intensity of
this band appears to be correlated with the intensity of the [NeII] at
12.8\,$\mu$m and of H$\alpha$, 
which shows that the emitters of the 11.3\,$\mu$m band are located
in the ionised gas.
The survival of AIB emitters in the HII region
could be due to the moderate intensity of the
radiation field (G$_0$\,$\sim$\,100) and the lack of photons with energy above $\sim$25\,eV.  
The enhancement of the
intensity of the 11.3\,$\mu$m band in the HII region, relative to
the other AIBs can be explained by the presence of neutral PAHs.}
{Our observations highlight a transition region
between ionised and neutral PAHs observed with ideal conditions in our Galaxy. A
scenario where PAHs can survive in HII regions and be significantly 
neutral could explain the detection of a prominent 11.3\,$\mu$m band
in other Spitzer observations.}


   \maketitle

\section{Introduction}

Polycyclic Aromatic Hydrocarbons (PAHs) were proposed twenty years ago
by \citet{Leger} and \citet{Allamandola} to explain the infrared
emission bands observed at 3.3, 6.2, 7.7, 8.6 and 11.3\,$\mu$m.  These
emitters are an ubiquitous component of interstellar dust which has
been observed with the Infrared Space Observatory (ISO) in a wide
range of interstellar conditions \citep[e.\,g.][]{Boulanger98a,
Uchida}.  These aromatic infrared bands (AIBs) have already been
observed in spectra attributed to HII regions
\citep[e.\,g.][]{Peeters, Hony, Vermeij} but never with clear proof
that their emitters are within the ionised gas rather than within an
associated photodissociation region on the same line of sight.
Moreover, several studies report the destruction of these emitters in
the HII regions of M17 and the Orion Bar.  In these regions, the
strong radiation field is thought to be the main cause of this
destruction \citep[e.\,g.][and reference therein]{Kassis}.  In this
paper, we use ``PAHs'' as a generic term in order to designate the
emitters of the AIBs although the exact nature of these emitters is
still a matter of debate.

\begin{figure}
   \centering
    \includegraphics[width=9cm]{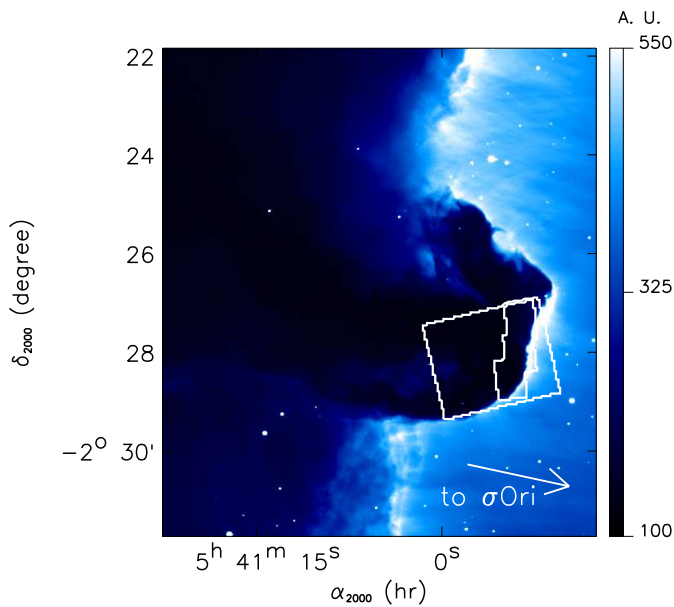}
    \caption{  H$\alpha$ map, in arbitrary units, obtained with the 0.9 m Kitt Peak National 
               Observatory (KPNO) telescope
               \citep{Pound}. The contours show the areas observed
               with both IRS-SL (small area) and IRS-LL (large area).}
    \label{fig:Halpha_map}
\end{figure}

\begin{figure}
   \centering
     \includegraphics[width=9cm]{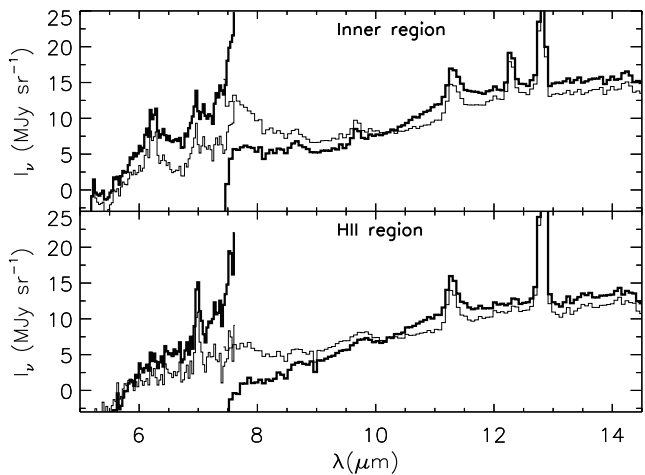}
      \caption{ Average SL spectra obtained within the Inner and HII
      regions (see the contours of these two regions in
      Fig.\,\ref{fig:multi_species}) before (thick line) and after
      (thin line) correction of the residual emission (see
      \S\,\ref{sect:observations}).}
         \label{fig:SL_correction}
\end{figure}

Dust grains can play an important role in the energetic balance of HII
regions through photoelectric heating \citep[and references
therein]{Weingartner}.  Since these processes are dominated by small
grains, the presence of PAHs in HII regions has a strong impact on the physics
of these objects.
 
In front of the western illuminated edge of the molecular cloud L1630,
the visible plates are dominated by extended red emission due to the
H$\alpha$ line emission emerging from the HII region IC434
\citep[e.\,g.][]{Louise}.  In the visible, the Horsehead nebula, also
known as B33 \citep{Barnard}, emerges from the edge of L1630 as a dark
cloud in the near side of IC434.  The Horsehead nebula is a familiar
object in astronomy and has been observed many times at visible,
IR and submm wavelengths \citep{Zhou, Abergel, Pound, Teyssier, Habart,
Pety, HilyB}.  IC434 and the Horsehead nebula are excited by the
$\sigma$ Orionis star which is an O9.5V binary system \citep{Warren}
with an effective temperature of $\sim$\,34\,600\,K
\citep{Schaerer}. L1630 is located at a distance of $\sim$400 pc\footnote{from
the study of the distances to B stars in the Orion association by
\citet{Anthony}.}.  Assuming that $\sigma$ Orionis and the Horsehead
are in the same plane perpendicular to the line of sight, the distance
between them is $\sim$\,3.5\,pc ($\sim$\,0.5\degr) which gives
G$_0$\,$\sim$\,100 \citep[energy density of the radiation field between 6 and
13.6 eV in unit of Habing field,][]{Habing} for the
radiation field which illuminates the Horsehead nebula.
  
In this paper, we study the AIBs from IC434 in front of the Horsehead
nebula observed with the Infrared Spectrograph \citep[IRS;][]{Houck}
on board the Spitzer Space Telescope \citep{Werner}.  The paper is
organised as follows : in \S\,\ref{sect:observations}, we present our
IRS data and the data processing. In \S\,\ref{sect:Horse_IRS}, we show
that IRS data confirm the description of the structure of the object
from previous studies. In the following section
(\S\,\ref{sect:HII_spectrum}), we extract the typical spectrum of the
HII region and study the location of the AIB emitters of this
spectrum.  We compare the HII region spectrum with the spectrum
obtained in the inner region in \S\,\ref{sect:comparaison}, and
propose a scenario to explain the observed spectral variations in
\S\,\ref{sect:aromatic}. The survival of PAHs in the HII region is
discussed in \S\,\ref{sect:survival}. We conclude in
\S\,\ref{sect:conclusion}.

\section{Observations \& data reduction}\label{sect:observations}

The Horsehead nebula has been observed with IRS as a part of our
``SPECPDR'' program \citep{Joblin2005} on 2004 October 2, and using
the Short-High (SH), Long-High (LH) , Short-Low (SL) and Long-Low (LL)
modules of the instrument. In this paper, we only present SL
(5.2-14.5\,$\mu$m, slit size: 57\arcsec\,$\times$\,3.6\arcsec,
R=64-128) and LL (14-38\,$\mu$m, slit size:
168\arcsec\,$\times$\,10.6\arcsec, R=64-128) observations.  We used
the ``spectral mapping mode''. An observation is made of
N$_{\rm{step}}$= 23 (SL) or 25 (LL) steps of half the slit width in
the direction perpendicular to the slit long axis.  For the SL module, three
observations were taken successively at three different positions on
the sky in order to perform a complete mapping of the illuminated edge
of the nebula.  The resulting observed areas are shown in
Fig.\,\ref{fig:Halpha_map}, overplotted on the H$\alpha$ map.  The
integration times were 14 and 60\,s per pointing for the second
(5.2-8.7\,$\mu$m) and the first (7.4-14.5\,$\mu$m) orders of SL,
respectively, and 14 s per pointing for both orders of LL.  

We have developed a pipeline which builds spectral cubes (two
spatial dimensions and one spectral dimension) in a homogeneous way
from the data (version S13) delivered by the Spitzer Science Center
(SSC). We start from the BCD level. One integration corresponds to
one BCD image. For each BCD image, we extract for all wavelengths an
image of the slit which is projected on the sky. For each observation,
made of N$_{\rm{step}}$ integrations, we build a spectral cube with
N$_{\rm{x}}$\,$\times$\,N$_{\rm{y}}$ spatial pixels and $\rm{N_w}$
spectral pixels. We keep the same wavelength sampling as in the BCD
images and the spatial grid has a pixel size of 2.5\arcsec\, for LL
and 0.9\arcsec\, for SL (which corresponds to half the pixel size
on the BCD images).  Whenever we study the full spectral range, we
also reproject the SL data on the LL grid.

\begin{figure*}
   \centering
    \includegraphics[width=6cm]{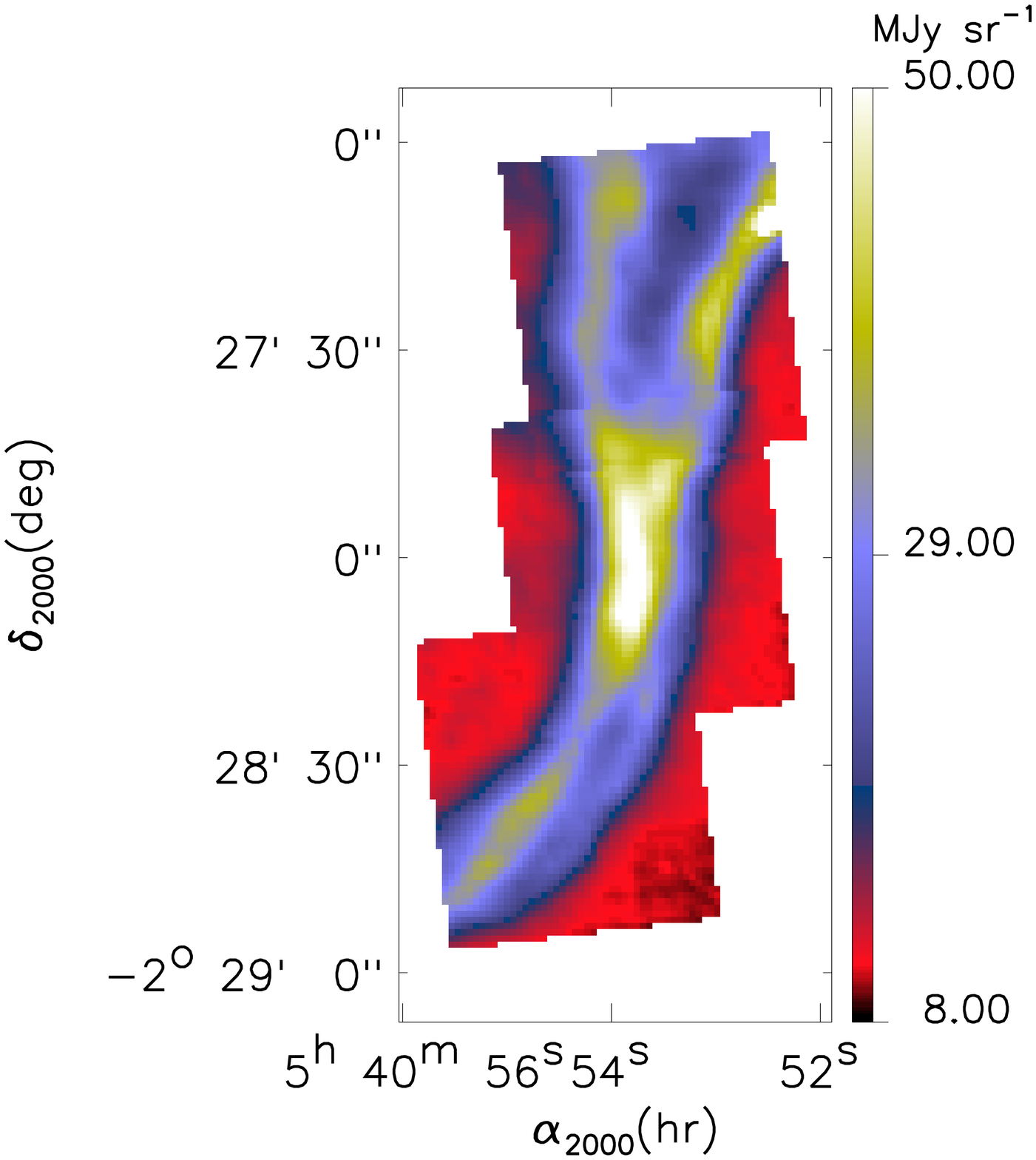}
    \includegraphics[width=11.5cm]{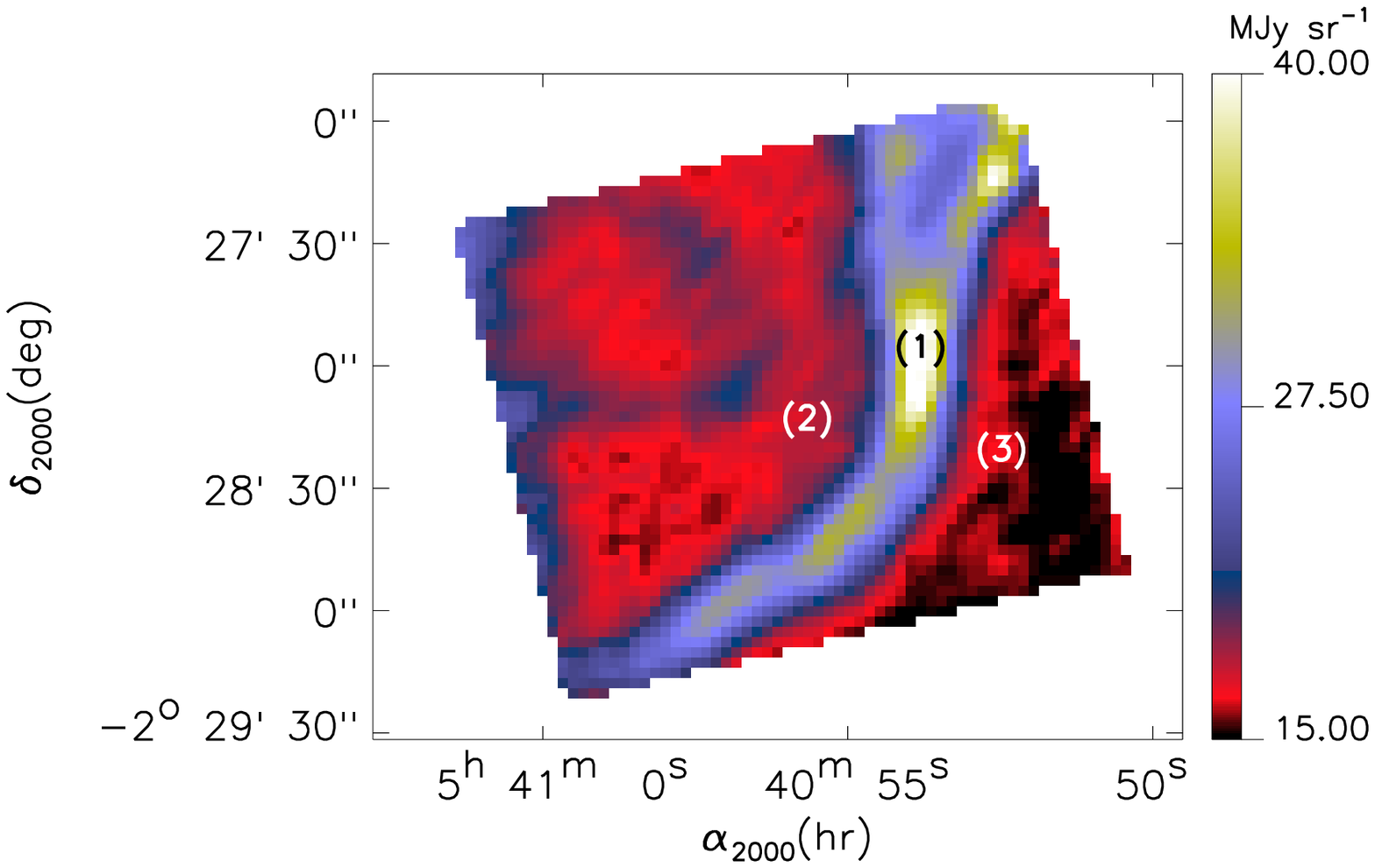}   
      \caption{Maps of the Horsehead nebula at
               11-11.5\,$\mu$m (SL, left panel) and 20-21\,$\mu$m (LL, right panel).
               The SL and LL maps have pixel size
               0.9\arcsec and 2.5\arcsec, respectively.  Positions of
               the spectra of Fig.\,\ref{fig:spectres_typiques} are
               shown on the LL map (right panel).}
         \label{fig:SL_LL_area}
\end{figure*}

We identify and correct the bad pixels not flagged out in the SSC
pipeline by median filtering on the combined N$_{\rm{step}}$ BCD images.
The data are flux-calibrated in Jansky using the pipeline S13 conversion factors
and the tuning coefficients given in the ``fluxcon'' table. Finally,
we derive extended emission flux intensities by using the Slit Loss
Correction Function due to the point-spread function overfilling the
IRS slit \citep{Smith}.  In the following, the LL data at
35-38\,$\mu$m are not considered due to the strong decrease of sensitivity.

Only the HII region in front the Horsehead nebula presents nearly flat
infrared emission and could have been used to define an {\it off}
spectrum.  Since the goal of this paper is precisely to study the
emission emerging from the HII region, we did not subtract any off
spectrum.
The lack of such correction explains the discontinuity
systematically found for all pixels between the first order and the
second order parts of the SL spectra, together with a systematic
decrease of the continuum with decreasing wavelength, down to negative
values for wavelengths below $\sim$6\,$\mu$m
(Fig.\,\ref{fig:SL_correction}). We find that for each pixel the
amplitude of these effects does not depend on the detected emission,
but appears strongly correlated with the non-zero emission detected in
the interorder regions of the BCD images which in principle does not
receive any incident photon. This non-zero emission does not depend on
the emission in the order regions of the BCD images, but presents some
correlation with the emission in the peakup region of the array.  It
could be a residual after the ``droop'' or the stray-light
corrections (see the IRS data handbook\footnote{see
http://ssc.spitzer.caltech.edu/irs/dh/}). We estimate the amplitude of
this residual in the order regions by extrapolating for each row the
emission detected in the interorder regions, and subtract this
residual from the BCD image. This subtraction is performed on the BCD
image multiplied by the flat field image (taken in the calibration
files delivered by the SSC), since we work in the hypothesis that the
effects we want to correct are additive. Then, we divide the corrected
BCD image by the flat field image. Finally, we build a corrected
spectral cube using the algorithm described above.  The correction
only affects the shape of the continuum emission and does not change
the amplitude of the spectral bands and lines.  The discontinuity
between the two orders of the SL spectra is strongly reduced
(Fig.\,\ref{fig:SL_correction}).  
However, we have to keep in mind
that at that time the origin of the corrected effects is not known,
therefore we must be cautious in the interpretation of the continuum
emission. 

\section{The Horsehead as seen by IRS}\label{sect:Horse_IRS}

\begin{figure*}
   \centering
    \includegraphics[width=12cm]{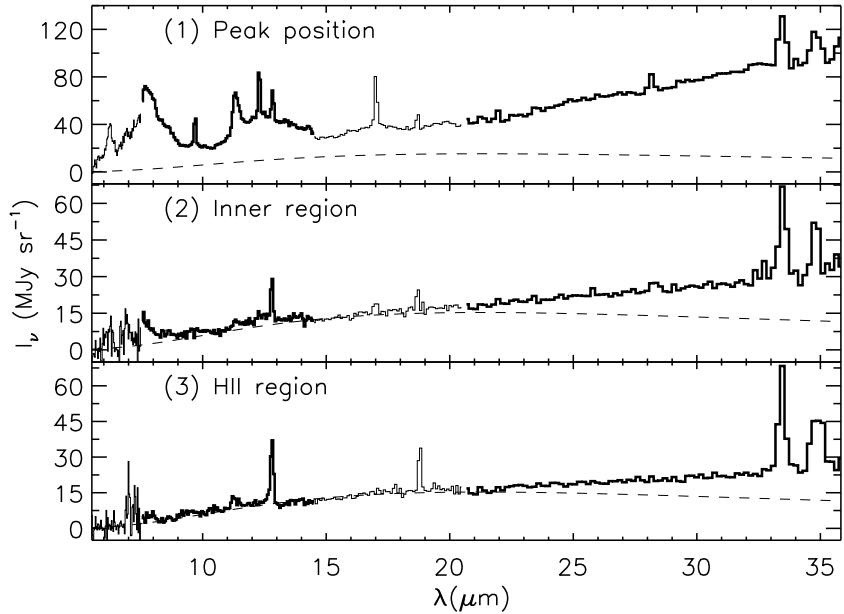}
      \caption{Typical spectra obtained by combining SL and LL data for individual
      pixels (1) at the infrared peak position, (2) in the inner
      region of the Horsehead nebula behind the emission peak
      and (3) in front of the dense cloud in the HII region, as 
      shown in the right panel of Fig.\,\ref{fig:SL_LL_area}.
      Alternating thicknesses correspond to SL2, SL1, LL2 and LL1 orders. The
      dashed lines show the contribution of the zodiacal emission (see
      \S\,\ref{sect:Horse_IRS}).  }
         \label{fig:spectres_typiques}
\end{figure*}

Figure \ref{fig:SL_LL_area} gives an example of the spectral maps
obtained from SL and LL observations. For all pixels within the SL
field, we have a full spectrum from 5 to 35\,$\mu$m
(Fig.\,\ref{fig:spectres_typiques}).
 
The zodiacal emission is computed using the SSC background
estimator\footnote{see
http://ssc.spitzer.caltech.edu/documents/background/} which is based
on the COBE/DIRBE model \citep{Kelsall}.  The dashed lines in
Fig.\,\ref{fig:spectres_typiques} show the zodiacal contribution to
our observations which is not simply the zodiacal emission at the date
and sky coordinates of our observations.  Indeed, the ``dark'' level
subtracted from the data in the SSC pipeline is obtained without
shutter by pointing a fixed area of the sky with faint infrared
emission (RA = 268$\,\fdg$96, DEC = 65$\,\fdg$43) as explained in the
IRS Data Handbook.  This ``dark'' level will
contain some zodiacal emission ($\sim$\,14\,MJy\,sr$^{-1}$ at
$\sim$\,18\,$\mu$m).  Therefore, the zodiacal contribution to our
observations is the difference between the zodiacal emission at the
time and the position of our observations and the zodiacal emission at
the ``dark'' position.  This zodiacal contribution does not vary
across the observed area of the sky and is accurate to
$\sim$14\%$^3$.

The spatial structure detected with IRS is comparable to the
broad-band observations taken with ISOCAM \citep{Abergel}, but we now
have the spectral information from 5 to 35\,$\mu$m and better spatial
resolution.  Thanks to the edge-on geometry of this PDR, it is
possible to perform spectral analysis of the emission in the HII
region, the edge of the PDR and the inner region inside the PDR,
separately.  Three illustrative spectra for individual pixels are
presented in Fig.\,\ref{fig:spectres_typiques}:
\begin{itemize}
\item[(1)] The first spectrum is taken at the infrared peak position
 and is typical for a PDR. It shows the main H$_2$ rotational lines
 (0-0\,S(4) to S(0) at 6.9, 9.7, 12.3, 17.0 and 28.2\,$\mu$m), the AIBs and continuum.
\item[(2)] The second spectrum is taken in the inner region behind the
 peak (to the east of the peak) and shows AIBs and H$_2$ emission
 lines with lower intensities since the emitting matter is located
 more deeply in the dense cloud.
\item[(3)] The third spectrum is taken in front of the illuminated
 surface in the HII region (to the west of the peak) and is dominated
 by fine structure lines of ionised species as expected for a
 HII region.  It shows the following lines: [ArII] at 6.98\,$\mu$m,
 [NeII] at 12.8\,$\mu$m, [SIII] at 18.7 and 33.4\,$\mu$m, [SiII] at
 34.8\,$\mu$m, but not the more excited lines: [NeIII] at
 15.5\,$\mu$m, [SIV] at 10.5\,$\mu$m and [ArIII] at 9.0\,$\mu$m.  It
 also contains the 11.3\,$\mu$m AIB.
\end{itemize} 

Spectra (1) and (2) also present ionised lines which are likely
emitted by the ionised medium surrounding the dense cloud.  The
continuum emission at wavelengths lower than $\sim$\,20\,$\mu$m
appears to be dominated by the zodiacal emission for the spectra (2)
and (3).

In the following, we remove the zodiacal contribution from all spectra
and focus our study on the SL spectra of the HII region which contains
the 11.3\,$\mu$m AIB.  The study of the spectral properties around the
peak position will be the subject of a forthcoming paper.

\section{HII region spectrum}\label{sect:HII_spectrum}

\begin{figure*}
   \centering
     \includegraphics[width=12cm]{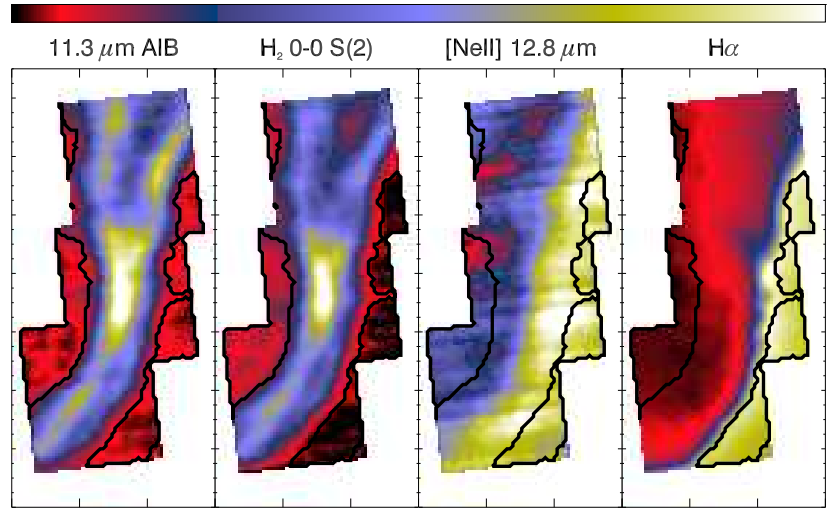}
     \caption{ Emission maps of the 11.3\,$\mu$m AIB, H$_2$\,0-0\,S(2),
               [NeII]\,at 12.8\,$\mu$m (as seen by IRS) and H$\alpha$
               \citep{Pound}.
               The contours show the area defined as
               the HII region
               (west of the peak) and the inner region (east of the
               peak) in \S\,\ref{sect:HII_spectrum} and
               \S\,\ref{sect:comparaison}.} 
         \label{fig:multi_species}
\end{figure*}
 
\begin{figure*}
   \centering
     \includegraphics[width=12cm]{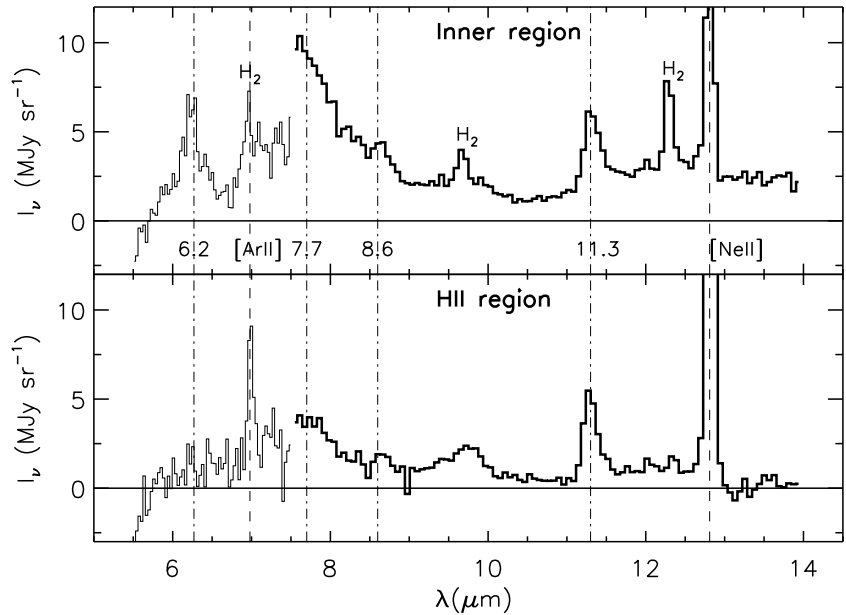}
     \caption{ Mean spectra of the HII region (lower) and the inner region (upper)
               as defined in \S\,\ref{sect:HII_spectrum} and
               \S\,\ref{sect:comparaison} and shown in Fig.\,\ref{fig:multi_species}.  
               The [NeII]
               line at 12.8\,$\mu$m is truncated for clarity.  In the
               HII region spectrum, the broad feature at
               $\sim$\,10\,$\mu$m is an artefact (see
               \S\,\ref{sect:HII_spectrum} and
               Fig.\,\ref{fig:spectres_HII_boite}).  The zodiacal
               emission has been removed from these spectra (see
               \S\,\ref{sect:Horse_IRS}).}
         \label{fig:spectre_compare}
\end{figure*}
 
Fig.\,\ref{fig:multi_species} shows emission
maps of the 11.3\,$\mu$m AIB \citep[fitted with a lorentzian profile
following][]{Boulanger98b}, H$_2$\,0-0\,S(2), [NeII] at 12.8\,$\mu$m
(both fitted with a gaussian profile) and H$\alpha$ obtained at the
Kitt Peak National Observatory (KPNO) telescope by \citet{Pound}
(Fig.\,\ref{fig:Halpha_map}).
Both [NeII] and H${\alpha}$ are emitted by ionised gas.  However, they
cannot be used alone to define the HII region and exclude PDR emission
which could be located on the same line of sight behind or in front of
the ionised gas (with respect to us).
We see in Fig.\,\ref{fig:multi_species} that the H${\alpha}$ and
[NeII] emissions do not peak at the same location due to projection
effects and the difference of extinction efficiency since they emit at
different wavelengths.
We need a PDR tracer in order to exclude PDR emission.  We use the
H$_2$\,0-0\,S(2) line at 12.3\,$\mu$m since it is a good PDR tracer
for this dense illuminated ridge \citep{Habart}, which is well
detected in our data.  Thus, we define the HII region as the area
where $\rm{I}_{\rm{H}{\alpha}}$\,$>$\,450 in arbitrary units and
I(H$_2$\,0-0\,S(2))\,$<$\,6\,10$^{-9}$\,W\,m$^{-2}$\,sr$^{-1}$
(detection limit due to the noise in individual spectra).  We use
H${\alpha}$ rather than [NeII] to trace the ionised gas since it is
not affected by ionisation fraction effects.  The contours of the HII
region we have defined are shown in Fig.\,\ref{fig:multi_species}.
The H${\alpha}$ emission in this HII region traces the ionised material
in front of the Horsehead.
In fact, the extinction due to this material on a line of sight is
A$\rm{_V}$\,$\sim$\,0.01-0.06, considering
$\rm{n_e}$\,$\sim$\,$\rm{n_H \sim 100-350}$\,cm$^{-3}$ (see appendix
\ref{app:ne}), a depth of $\sim$0.1\,pc for the Horsehead nebula
\citep{Habart} and
$\rm{N_H/A_V}$\,$\sim$\,$\rm{1.87\,10^{21}\,cm^{-2}\,mag^{-1}}$
\citep{Bohlin78}.

\begin{figure*}
   \centering
   \includegraphics[width=4.5cm]{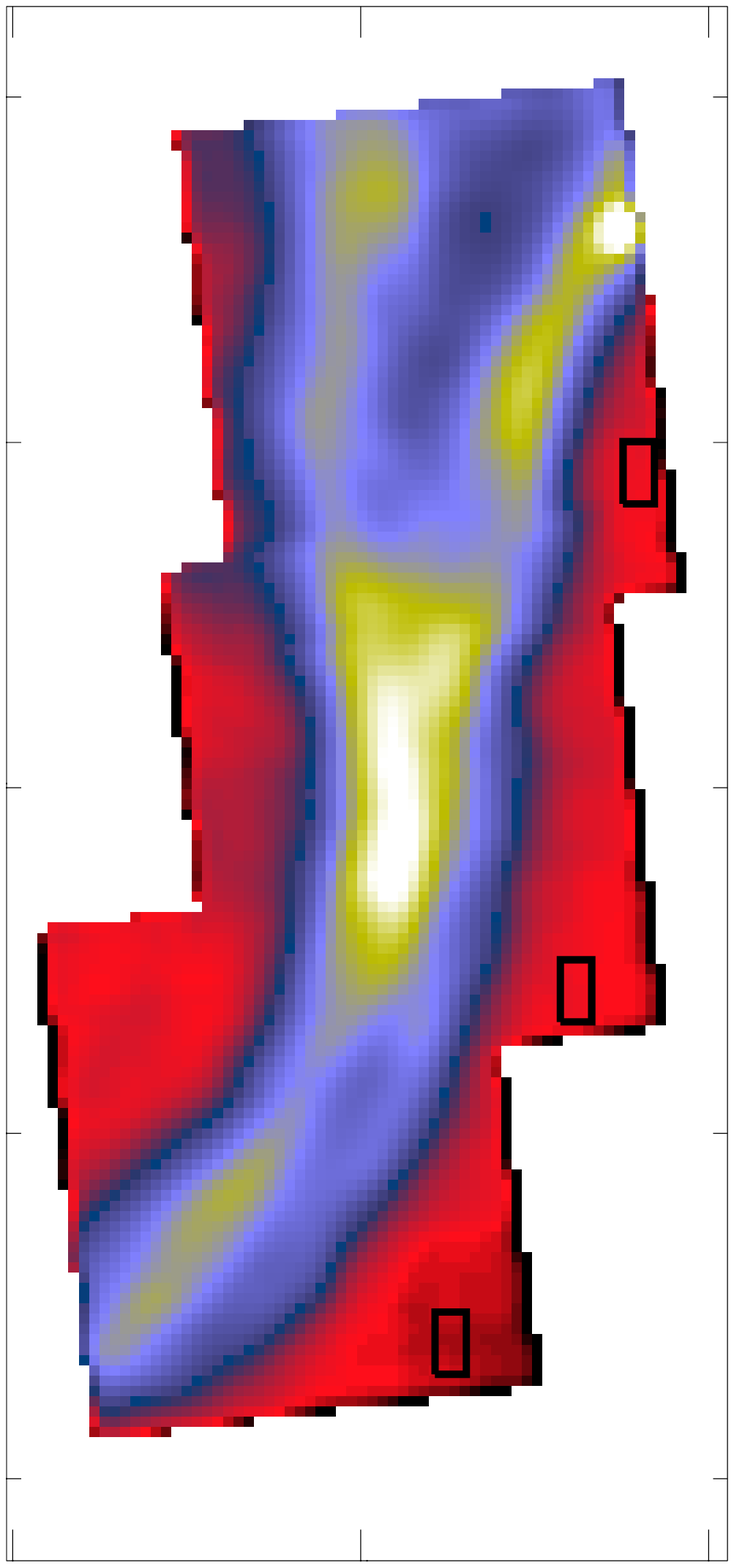}
   \includegraphics[width=10.cm]{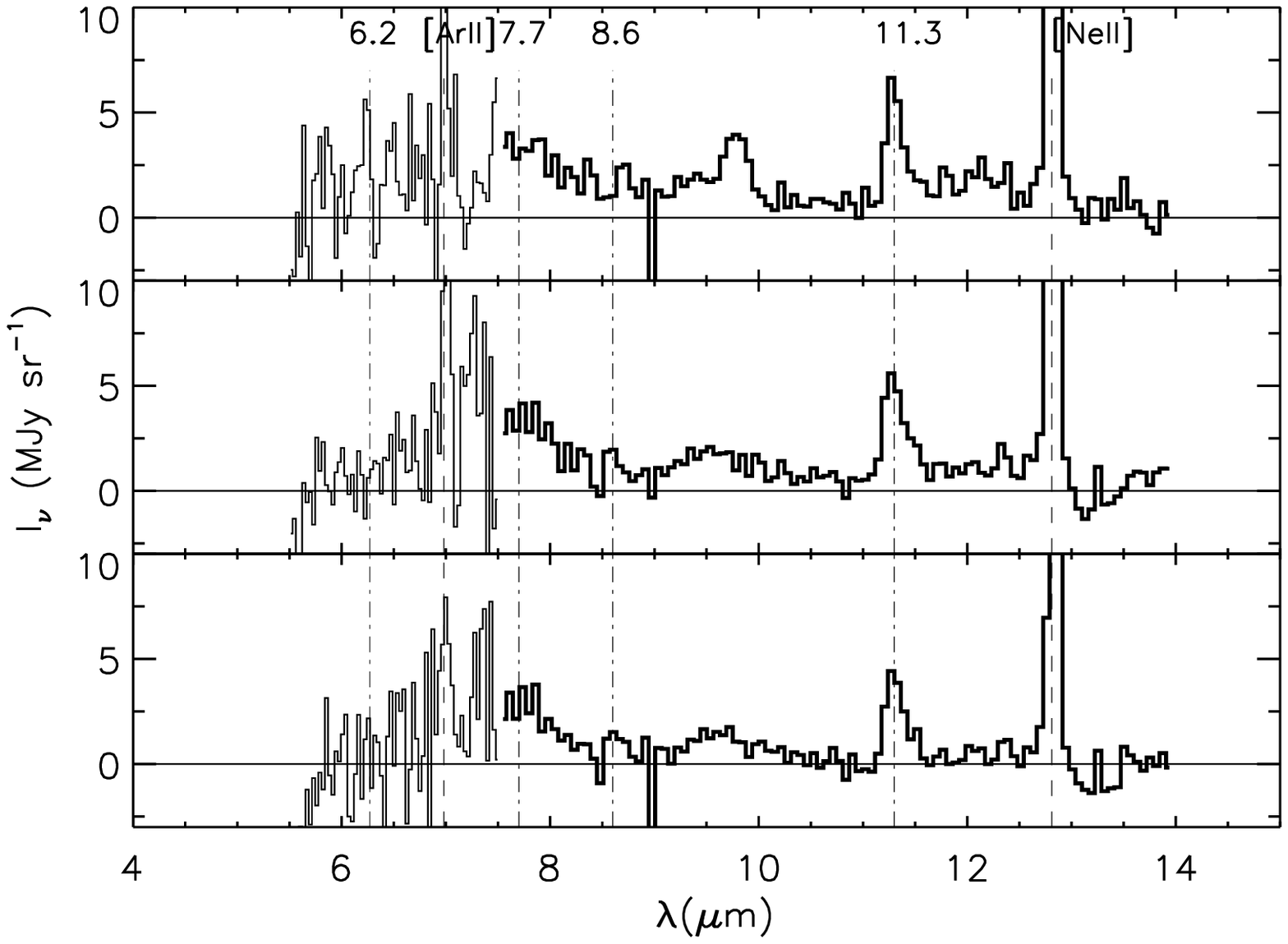}
      \caption{ Spectra from three areas (small black rectangles)
                shown on the 11-11.5\,$\mu$m map (top, middle and lower
                panels correspond to the north, the central and the
                south area, respectively). The zodiacal emission has
                been subtracted (see \S\,\ref{sect:Horse_IRS}). For
                clarity, the [NeII] line is truncated.
                The broad feature at $\sim$\,10\,$\mu$m in the spectrum 
                of the top panel 
                is an artefact (see \S\,\ref{sect:HII_spectrum}).}
         \label{fig:spectres_HII_boite}
\end{figure*}

The average spectrum computed within this area
(Fig.\,\ref{fig:spectre_compare}, lower spectrum) shows the [ArII] and [NeII] lines at
6.98\,$\mu$m and 12.8\,$\mu$m, respectively.  A band is also detected
at 11.3\,$\mu$m with a surprisingly high intensity compared to the
intensities of other AIBs at 6.2, 7.7 and 8.6\,$\mu$m.  Other bands
seem necessary to account for the emission plateau between 11.3 and
13\,$\mu$m.  A broad emission feature is visible around 10\,$\mu$m,
however its amplitude is variable (Fig.\,\ref{fig:spectres_HII_boite})
and depends strongly on the offset position from the center of the
slit (i.e. on the position on the detector).  We therefore
conclude that this feature is mainly due to an artefact.  On the
other hand, the AIB at 11.3\,$\mu$m is real since it appears
everywhere in the HII region with a strong amplitude, as shown in
Fig.\,\ref{fig:spectres_HII_boite}.

\begin{figure}
   \centering
   \includegraphics[width=9.cm]{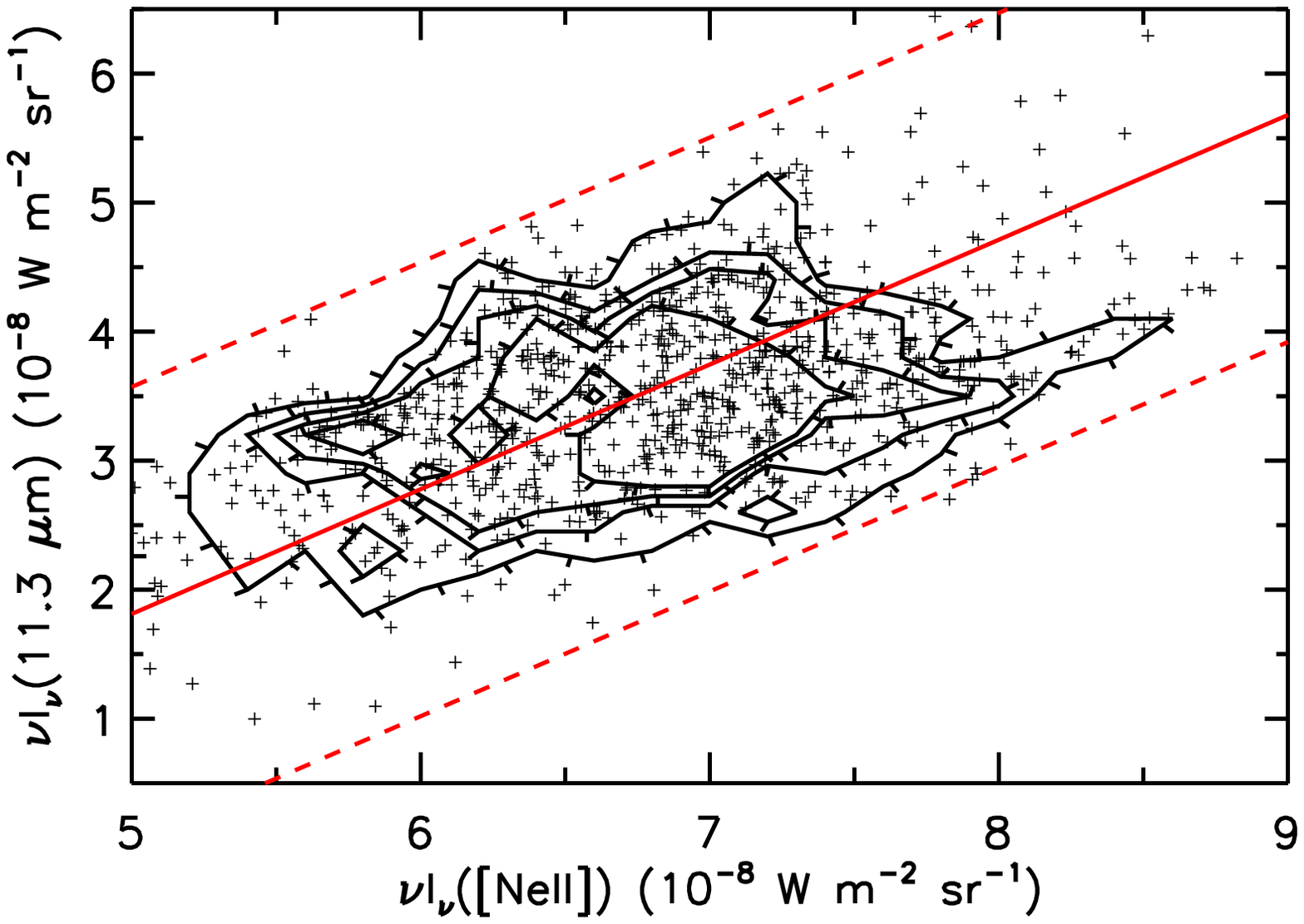}
   \includegraphics[width=9.cm]{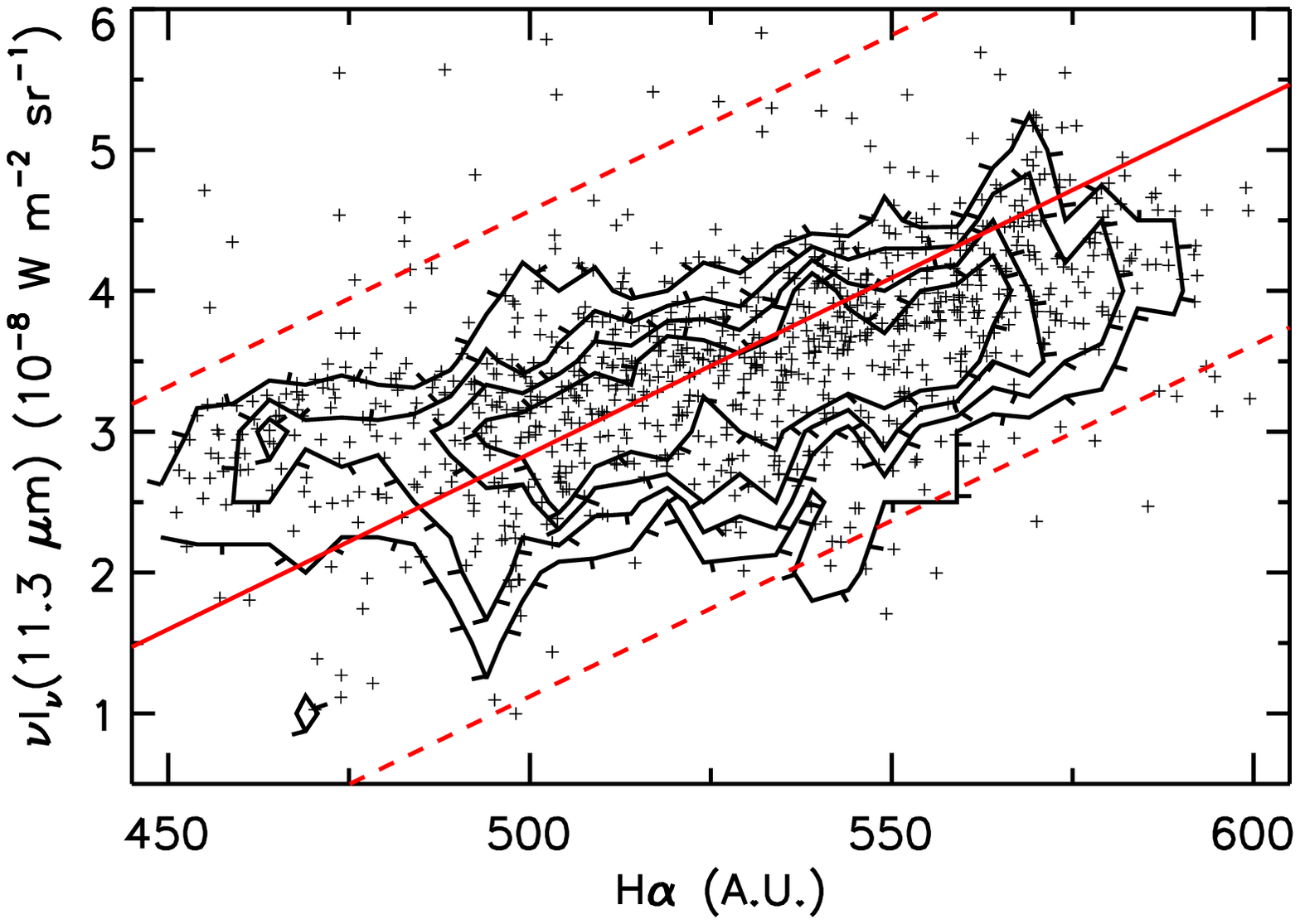}
     \caption{Relationships between the [NeII], H$\alpha$ 
              and $11.3\,\mu \rm{m}$
              for pixels of the HII region
              (defined in \S\,\ref{sect:HII_spectrum} and shown in
              Fig.\,\ref{fig:multi_species}). Contours are
              histograms of the point densities.
              The red straight lines
              are linear fits to the data.
              The red dashed lines are the 3$\sigma$ limit
              with respect to the linear fitting.
              }
         \label{fig:correlation}
\end{figure}

A method to check whether the 11.3\,$\mu$m emitters are located in the
ionised gas is to look for spatial correlation between the
11.3\,$\mu$m AIB and the ionised gas lines emission in the HII
region.  
The figure\,\ref{fig:correlation} presents the $11.3\,\mu$m AIB vs
[NeII] and vs H$\alpha$ correlation diagrams for pixels of the HII
region.  
The computed [NeII] line intensity 
is not affected by a possible contribution
of the 12.7\,$\mu$m AIB, since 
our spectral decomposition takes into account this AIB.
We see that the $11.3\,\mu$m AIB intensity is correlated with both the
[NeII] and H$\alpha$ ones which shows that the observed correlation is
not dominated by systematic effects in the IRS data.  
The correlation coefficients are 0.56 and 0.63 for $11.3\,\mu$m
AIB vs [NeII] and $11.3\,\mu$m AIB vs H$\alpha$, respectively.  Note
that we have only considered the points within the 3$\sigma$ limit
with respect to the linear fitting (dashed lines in
Fig.\,\ref{fig:correlation}). The relatively low values of the
correlation coefficients are due to 
systematic effects on the IRS computed maps (striping,
Fig.\,\ref{fig:multi_species}) and to the fact that the relationship
between the ionised gas lines intensity and the AIBs intensity is
not systematically linear.

\begin{figure}
   \centering
     \includegraphics[width=9cm]{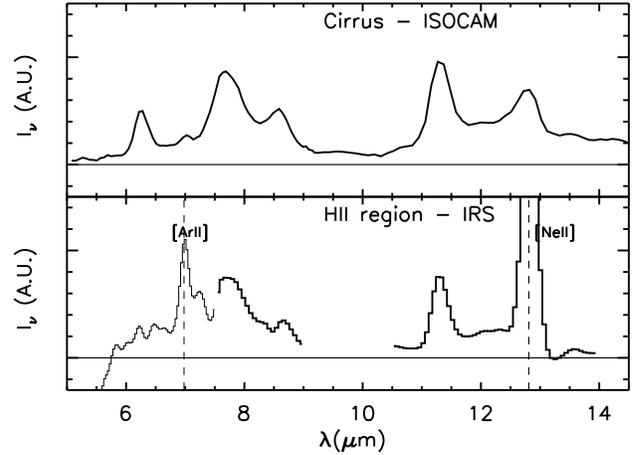}
      \caption{Cirrus spectrum observed with ISOCAM \citep{Flagey} and HII region
               spectrum as defined in \S\,\ref{sect:HII_spectrum}
               convolved with a gaussian yielding a resolving power 
               of 40, comparable to ISOCAM-CVF.
               The 9-10.5\,$\mu$m range is suppressed since it contain
               an artefact (\S\,\ref{sect:HII_spectrum}).  }
         \label{fig:compare_HII_cirrus}
\end{figure}

At least about half the 11.3\,$\mu$m emitters on the line of sight
should be located in the ionised gas in front of the Horsehead nebula
where the H$\alpha$ and [NeII] emissions rise, since the two
correlations range from $\sim$\,2.5 to
$\sim$\,4.5\,10$^{-8}$\,W\,m$^{-2}$\,sr$^{-1}$ for the 11.3\,$\mu$m
AIB intensity.  
A typical Cirrus spectrum has been estimated by \citet{Flagey} from
ISOCAM data (Fig.\,\ref{fig:compare_HII_cirrus}, upper spectrum). It
presents a shape comparable to the HII spectrum, except for the
6.2\,$\mu$m AIB which indicates that the contribution of Cirrus
emission does not dominate the detected HII spectrum.  In any case,
the contribution of Cirrus should be comparable for the HII region and
inner region spectra (as defined in the next section), and cannot
affect our conclusions based on spectral variations between these two
spectra.

\section{Comparison with the inner region spectrum}\label{sect:comparaison}

As for the HII region (\S\,\ref{sect:HII_spectrum}), we
define an inner region where $\rm{I}_{\rm{H}{\alpha}}$\,$<$\,135 in
arbitrary units and $\rm{I}_{\nu,11-11.5\mu\rm{m}}$\,$<$\,15
MJy\,sr$^{-1}$ in order to avoid the bright infrared filament
(Fig.\,\ref{fig:multi_species}) were the [NeII] emission peaks
due to projection effets.  The average spectrum computed within this
area (Fig.\,\ref{fig:spectre_compare}, upper spectrum) shows the
0-0\,S(3)-9.7\,$\mu$m and 0-0\,S(2)-12.3\,$\mu$m H$_2$ rotational
lines, the AIBs at 6.2, 7.7, 8.6 and 11.3\,$\mu$m and the [NeII] line
at 12.8\,$\mu$m.  The 0-0\,S(5)-6.9\,$\mu$m H$_2$ and [ArII]
(6.98\,$\mu$m) lines are blended.

The presence of [NeII] and [ArII] lines indicates that this spectrum
contains ionised gas emission and could contain AIBs emitted in the
background and foreground ionised medium.
However, such a contribution from HII region AIBs will not change the conclusions of
our analysis of the spectral variations that we study in the
following.
  
The relative intensity of the AIBs presents striking differences
between the HII region and inner region spectra
(Fig.\,\ref{fig:spectre_compare}). While the intensity of the 11.3\,$\mu$m band is
comparable, the 6.2, 7.7 and 8.6\,$\mu$m bands present a spectacular
decrease in the HII region (more than a factor 2-3).  
The 6.2 and 7.7\,$\mu$m bands are
attributed to C-C stretching mode, the 8.6\,$\mu$m band to in-plane
C-H bending modes, and the 11.3\,$\mu$m band to C-H out-of-plane
bending mode.
In the following
section, we study the different processes which could explain such
spectral variation of the 6-9\,$\mu$m\,/\,11.3\,$\mu$m ratio.

\section{Interpretation of the spectral properties}\label{sect:aromatic}

\subsection{Hydrogenation state}

Hydrogenation state of PAHs can be traced by relative intensities of
the 11.3, 12, 12.7 and 13.6\,$\mu$m bands \citep[e.\,g.][]{Schutte}
which are C-H out-of-plane bending modes with one, two, three and four
H atoms on the same aromatic ring, respectively
\citep[e.\,g.][]{Leger, Hony}.  Both inner region and HII region
spectra (Fig.\,\ref{fig:spectre_compare}) present an 11.3\,$\mu$m band
and an emission plateau between 11.3 and 13\,$\mu$m above the
continuum, which can be attributed to hydrogenated PAHs \citep[high
hydrogenation coverage of PAHs has already been reported in HII
regions and PDRs, e.\,g.][]{Vermeij, Hony}.  Moreover the intensity of
the 11.3\,$\mu$m band compared to the plateau does not present any
strong variation between the two spectra, which indicates that the
hydrogenation states are comparable. We conclude that hydrogenation
effects are likely not the main process which could explain the
difference in the 6-9\,$\mu$m\,/\,11.3\,$\mu$m ratio between the two
spectra.

\subsection{Size distribution}

Using the model of \citet{Verstraete}, we find that the emission
ratio of 6-9\,$\mu$m\,/\,11.3\,$\mu$m is reduced by a factor of 3 only
if the size distribution contains exclusively PAHs bigger than 10$^3$ C atoms
($\sim$\,30\,\AA) in the HII region while it is a classical size
distribution \citep[mean size\,$\sim$\,6\,\AA, e.g.][]{Bakes3} in the inner
region. Then, we conclude that a change in the size distribution due
to destruction of smallest species cannot explained the
6-9\,$\mu$m\,/\,11.3\,$\mu$m ratio variation.

\subsection{Charge state}

Theoretical \citep[e.\,g.][]{Langhoff, Bakes,
Bakes2, Bauschlicher} and experimental \citep[e.\,g.][]{Szczepanski}
works show that the charge state of PAHs has a strong impact on the
6-9\,$\mu$m\,/\,11.3\,$\mu$m ratio.  Neutral PAHs emit significantly
less at 6-9 $\mu$m than at 11-13 $\mu$m with respect to charged ones (both anions and cations).  The
inner region spectrum which comes from neutral gas presents a high
value of the 6-9\,$\mu$m\,/\,11.3\,$\mu$m ratio, which can be
explained by charged PAHs (anions or cations). On the contrary, the
low value of the 6-9\,$\mu$m\,/\,11.3\,$\mu$m ratio in the HII region
spectrum can be explained by the presence of neutral PAHs.  The
spectra extracted from ISOCAM observations (4-16\,$\mu$m) of NGC7023
present comparable relative intensity variations attributed to charge effects
\citep{Rapacioli}.  
Our spectra from the HII region and the inner
region (Fig.\,\ref{fig:spectre_compare}) could therefore be 
attributed to PAH$^0$ and PAH$^+$, respectively.
 
The charge state of PAHs is mainly determined by the balance between
photoionisation and recombination rates of electrons
\citep{Weingartner, Bakes3} which is generally described by the ratio
of the UV intensity to the electronic density, G$_0$/n$_{\rm{e}}$.
The presence of positively charged PAHs in the inner region can be
explained by (1) the presence of UV photons which efficiently ionise
the PAHs and (2) a lack of free electrons for the recombination
(n$_{\rm{e}}$/n$_{\rm{H}}$\,$\sim$\,[C]/[H]\,$\sim$\,10$^{-4}$, C$^+$
being the main provider of electrons in the PDR).
 
For the HII region, we use version 05.07 of Cloudy \citep{Ferland} in order to
derive a quantitative estimate of the charge state of PAHs
\citep{vanHoof, Weingartner} in a fully
(n$_{\rm{e}}$\,$\sim$\,n$_{\rm{H}}$) ionised medium.  We perform a
simple model with an incident radiation field defined for an O9.5V
star of the Costar catalogue \citep{Schaerer} and located at 3.5
pc. The gas is taken to be at T\,=\,7500\,K \citep{Ferland1} and with
n$_{\rm{e}}$\,$\sim$\,100-350\,cm$^{-3}$ (see appendix
\ref{app:ne}). For PAHs with radius from 4.5 to 10.5\,$\AA$ and
distributed as n(a)\,$\propto$\,a$^{-3.5}$ \citep{Bakes3}, we obtain a
mean charge of 0.55\,-\,0.75 electron per PAH, corresponding to a
fraction of neutral PAHs in the HII region of 25\,-\,45\%.  Moreover,
the 6-9\,$\mu$m\,/\,11.3\,$\mu$m ratio of the HII region spectrum is
in agreement with those predicted by the emission model of
\citet{Bakes} for such a charge distribution.  We conclude that the
HII region spectrum can be explained by a mixture of neutral and
anionic PAHs.

\section{PAHs survival in the HII region}\label{sect:survival}

The 11.3\,$\mu$m band is observed in the HII region up to a distance
of $\sim$\,20\,\arcsec (or 0.04\,pc) from the ionisation front
(see for instance Fig.\,\ref{fig:multi_species}).  This distance can be translated to a
lower limit of the survival time of the emitters equal to
$\sim$\,5\,10$^3$ years when considering that the gas just ionised
at the ionisation front expands freely in the HII region at the sound
speed c$_{\rm{s}}$\,=\,$\sqrt{\frac{\gamma \rm{k T}}{\mu
\rm{m_H}}}$\,$\sim$\,10\,km\,s$^{-1}$ for T\,=\,7500\,K, $\mu$\,=\,0.7
and $\gamma$\,=\,5/3.
 
We have seen in \S\,\ref{sect:Horse_IRS} that in the HII region, all
species with an ionisation potential (IP) lower or equal than that of
[SIII] (IP=23.34\,eV) are detected while species with IP higher or
equal than that of [ArIII] (IP=27.63\,eV) are not detected.  Thus, the
incident radiation field in the HII region contains UV photons with
energy up to $\sim$\,25\,eV.
Thus, some aromatic emitters can survive to a radiation field with
G$_0$\,$\sim$\,100 and photons up to $\sim$\,25\,eV. For comparison,
in more highly excited HII regions such as the Orion Bar, the radiation field
is more intense (G$_0$ larger than 10$^4$) and also harder
since the exciting star is an O6
\citep[$\rm{\theta^1\,Ori\,C}$, e.g.][]{Kassis, Allers} 
with $\rm{T_{eff}}\,\sim\,44000\,K$.
There, PAHs are
destroyed on time scales lower than 1000\,years \citep{Kassis}.

It must be emphasized that the presence of AIB emitters in the
ionised gas could be related to the continuous injection of 
``fresh'' matter due to photoevaporation of the Horsehead.

\section{Conclusion}\label{sect:conclusion}

Our main observational result is the detection of a strong
11.3\,$\mu$m emission band in the HII region facing the Horsehead
nebula. The spectral imaging capabilities of IRS allow us to show that
the integrated intensity of the $11.3\,\mu \rm{m}$ AIB is correlated
with those of the [NeII] line and H$\alpha$ in the HII region.
Moreover the spectral variations of the AIB spectrum is clearly
spatially correlated with the variations of physical conditions
between the PDR and the HII region. Thus, to our knowledge, {\it this
is the first time that we detect the presence of AIB emitters in
ionised gas}.  Consequently, the $11.3\,\mu \rm{m}$ emitters are not
efficiently destroyed by the incident UV photons which have an energy
below $\sim$\,25\,eV as suggested by the detected ionised species.
The survival of PAHs in the HII region could be due to the moderate
intensity of the radiation field (G$_0$\,$\sim$\,100) and the lack of
photons above $\sim$\,25\,eV, compared to more highly excited HII
regions (in terms of intensity and hardness), where PAHs can be
destroyed on time scales lower than 1000\,years \citep{Kassis}.  It
could also be related to the continuous photoevaporation of the
Horsehead nebula which bring ``fresh'' matter into the ionised gas.
   
The enhancement of the intensity of the 11.3\,$\mu$m band in the HII
region, relative to the other AIBs, can be explained by the presence
of neutral PAHs. Our modelling of the charge state of PAHs with Cloudy
confirms that the HII region should contain a significant amount of
neutral PAHs. On the contrary, PAHs from the inner region must be
positively charged.  Variations of the size distribution of PAHs could
also affect the 6-9\,$\mu$m\,/\,11.3\,$\mu$m ratio
\citep[e.\,g.][]{Verstraete} but with a lower amplitude than
charge state variations.
 
In galaxies, the presence of neutral PAHs has been suggested by
\citet{Kaneda} to explain the prominent emission feature at
11.3\,$\mu$m compared to the 6.2, 7.7 and 8.6\,$\mu$m features
observed with Spitzer in several elliptical galaxies.
The present IRS observations of the Horsehead nebula thus provide a
textbook example in our Galaxy of the transition region between
ionised and neutral PAHs and allow to derive a physical scenario in order
to interpret extra-galactic spectra.
 
\appendix

\section{Determination of the electronic density}\label{app:ne}

The electronic density in the HII region is estimated using the
intensity ratio of the [SIII] lines at 18.7 and 33.4\,$\mu$m. These
two lines are fitted with gaussian profiles in order to compute their
integrated intensities for all pixels in the HII region (see
\S\,\ref{sect:HII_spectrum} and Fig.\,\ref{fig:multi_species}). The
mean intensities are I([SIII],
19\,$\mu$m)\,=\,2.69\,$\pm$\,0.11\,10$^{-8}$\,W\,m$^{-2}$\,sr$^{-1}$
and I([SIII],
33\,$\mu$m)\,=\,3.91\,$\pm$\,0.42\,10$^{-8}$\,W\,m$^{-2}$\,sr$^{-1}$.
Error bars include the calibration uncertainties (absolute stellar
model (1\,$\sigma$\,=\,3\%), the repeatability ($\sim$\,3\%) and the
uncertainty of the calibration at 33.4\,$\mu$m ($\sim$\,10\%)).  We
have I([SIII], 19\,$\mu$m)\,/\,I([SIII],
33\,$\mu$m)\,=\,0.69\,$\pm$\,0.08.  Consequently, using atomic
constants from \citet{Galavis} and \citet{Mendoza} and assuming a gas
temperature of 7500\,K \citep{Ferland1}, we derive an electronic
density n$_{\rm{e}}$\,$\sim$\,100-350\,cm$^{-3}$.


\begin{thebibliography}{}

  \bibitem[Abergel et al.(2003)]{Abergel}
    Abergel, A., Teyssier, D., Bernard, J.P. et al., 2003, A\&A, 410, 577
  \bibitem[Allamandola et al.(1985)]{Allamandola} 
    Allamandola, L. J., Tielens, A. G. G. M. \& Barker, J. R., 1985, ApJL, 290, L25
  \bibitem[Allers et al.(2005)]{Allers}
    Allers, K. N., Jaffe, D. T., Lacy, J. H et al., 2005, ApJ, 630, 368
  \bibitem[Anthony-Twarog(1982)]{Anthony}
    Anthony-Twarog, B. J.,  1982, AJ, 87, 1213
  \bibitem[Bakes et al.(2001a)]{Bakes}
    Bakes, E. L. O., Tielens, A. G. G. M., Bauschlicher, Charles W., Jr., 2001a, ApJ, 556, 501
  \bibitem[Bakes et al.(2001b)]{Bakes2}
    Bakes, E. L. O., Tielens, A. G. G. M., Bauschlicher, Charles W., Jr. et al., 2001b, ApJ, 560, 261
  \bibitem[Bakes \& Tielens(1994)]{Bakes3}
    Bakes, E. L. O. \& Tielens, A. G. G. M., 1994, ApJ, 427, 822
  \bibitem[Barnard(1919)]{Barnard}
    Barnard, E. E., 1919, ApJ, 49, 1
  \bibitem[Bauschlicher(2002)]{Bauschlicher}
    Bauschlicher, Charles W., Jr., 2002, ApJ, 564, 782	
  \bibitem[Bohlin et al.(1978)]{Bohlin78}
    Bohlin, R. C., Savage, B. D., Drake, J. F., 1978, ApJ, 224, 132-142
  \bibitem[Boulanger et al.(1998a)]{Boulanger98a}  
    Boulanger, F., Abergel, A., Bernard, J. P. et al., 1998a, ASPC, 132, 15
  \bibitem[Boulanger et al.(1998b)]{Boulanger98b}
    Boulanger, F., Boisssel, P., Cesarsky, D. et al., 1998b, A\&A, 339, 194 
  \bibitem[Ferland(2003)]{Ferland1}
    Ferland, G. J., 2003, ARA\&A, 41, 517
  \bibitem[Ferland et al.(1998)]{Ferland}
    Ferland, G. J., Korista, K. T., Verner, D. A. et al., 1998, PASP, 110, 761
  \bibitem[Flagey et al.(2006)]{Flagey}   
    Flagey, N., Boulanger, F., Verstraete, L. et al., 2006, A\&A, 453, 969   
  \bibitem[Galavis et al.(1995)]{Galavis}
    Galavis, M. E., Mendoza, C., Zeippen, C. J.,  1995, A\&AS, 111, 347
  \bibitem[Habart et al.(2005)]{Habart}
    Habart, E. , Abergel, A., Walmsley, C. et al., 2005, A\&A, 437, 177
  \bibitem[Habing(1968)]{Habing}
    Habing, H. J.,  1968, Bull. Astr. Netherlands, 19, 421
  \bibitem[Hily-Blant et al.(2005)]{HilyB}
    Hily-Blant, P., Teyssier, D., Philipp, S. et al., 2005, A\&A, 440, 909
  \bibitem[Hony et al.(2001)]{Hony}
    Hony, S., Van Kerckhoven, C., Peeters, E. et al., 2001, A\&A, 370, 1030
  \bibitem[Houck et al.(2004)]{Houck}
    Houck, J. R., Roellig, T. L., van Cleve, J. et al., 2004, ApJS, 154, 18 
   \bibitem[Joblin et al.(2005)]{Joblin2005}
    Joblin, C., Abergel, A., Bernard, J.-P. et al., 2005, IAUS, 231, 153
  \bibitem[Kaneda et al.(2005)]{Kaneda}
    Kaneda, H., Onaka, T., Sakon, I., 2005, ApJ, 632L, 83
  \bibitem[Kassis et al.(2006)]{Kassis}
    Kassis, Marc, Adams, Joseph D., Campbell, Murray F., 2006, ApJ, 637, 823 
  \bibitem[Kelsall et al.(1998)]{Kelsall}
    Kelsall, T., Weiland, J. L., Franz, B. A. et al., 1998, ApJ, 508, 44
  \bibitem[Langhoff(1996)]{Langhoff}
    Langhoff, S. R, 1996, J. Phys. Chem., 100, 2819
  \bibitem[L\'eger \& Puget(1984)]{Leger} 
    L\'eger A. \& Puget J.L., 1984, A\&A, 137, L5
  \bibitem[Louise(1982)]{Louise}
    Louise, R., 1982, Ap\&SS, 85, 405
  \bibitem[Peeters et al.(2002)]{Peeters}
    Peeters E., Mart\'in-Hern\'andez N. L., Damour F. et al., 2002, A\&A, 381, 571
  \bibitem[Pety et al.(2005)]{Pety}
    Pety, J., Teyssier, D., Foss\'e, D. et al., 2005, A\&A, 435, 885
  \bibitem[Mendoza \&  Zeippen(1982)]{Mendoza}
    Mendoza, C.,  Zeippen, C. J.,  1982, MNRAS, 199, 1025
  \bibitem[Pound et al.(2003)]{Pound}
    Pound, Marc W., Reipurth, Bo, Bally, John, 2003, AJ, 125, 2108
  \bibitem[Rapacioli et al.(2005)]{Rapacioli}
    Rapacioli, M., Joblin, C., Boissel, P., 2005, A\&A, 429, 193 
  \bibitem[Schaerer \& de Koter(1997)]{Schaerer}
    Schaerer, D. \& de Koter, A., 1997,A\&A, 322, 598
  \bibitem[Schutte et al.(1993)]{Schutte}
    Schutte, W. A., Tielens, A. G. G. M., Allamandola, L. J., 1993, ApJ, 415, 397
  \bibitem[Smith et al.(2007)]{Smith}
    Smith, J.D., Draine, B.T., Dale, D.A. et al., 2007, ApJ, 656, 770-791
  \bibitem[Szczepanski et Vala(1993)]{Szczepanski}
    Szczepanski Jan et Vala Martin, 1993, ApJ, 414, 646
  \bibitem[Teyssier et al.(2004)]{Teyssier}
    Teyssier, D., Foss\'e, D., Gerin, M. et al., 2004, A\&A, 417 ,135
  \bibitem[Uchida et al.(2000)]{Uchida}
    Uchida, K. I., Sellgren, K., Werner, M. W. et al., 2000, ApJ, 530, 817
  \bibitem[van Hoof et al.(2004)]{vanHoof}
    van Hoof, P. A. M., Weingartner, J. C., Martin, P. G. et al, 2004, MNRAS, 350, 1330
  \bibitem[Vermeij et al.(2002)]{Vermeij}
    Vermeij, R., Peeters, E., Tielens, A. G. G. M. et al.,  2002, A\&A, 382, 1042
  \bibitem[Verstraete et al.(2001)]{Verstraete}
    Verstraete, L., Pech, C., Moutou, C. et al., 2001, A\&A, 372, 981
  \bibitem[Warren \& Hesser(1977)]{Warren}
    Warren, W. H., Jr.\& Hesser, J. E., 1977, ApJS, 34, 115 
  \bibitem[Weingartner \& Draine(2001)]{Weingartner}
    Weingartner, Joseph C. \& Draine, B. T., 2001, ApJS, 134, 263
  \bibitem[Werner et al.(2004)]{Werner}
    Werner, M. W., Roellig, T. L., Low, F. J. et al, 2004, ApJS, 154, 1
  \bibitem[Zhou et al.(1993)]{Zhou}
    Zhou, S., Jaffe, D. T., Howe, J. E. et al., 1993, ApJ, 419, 190

\end{thebibliography}
\end{document}